\def\siN{\sum\limits^N_{i=1}}

\def\lll{\left[}
\def\rr{\right]}
\def\LL{\left(}
\def\RR{\right)}
\def\dspls#1{\displaystyle#1}

\documentstyle{article}
\begin{document}

\begin{center}
  A Model for \\
  Double-Stranded Excitations of DNA  \vspace{15mm}  \\
  V.L. Golo$^{+}$, E.I. Kats$^{++}$,
  and Yu.M. Yevdokimov$^{+++}$ \vspace{5mm} \\
  $^{+}$Moscow University, Department of Mechanics and Mathematics,
  Moscow, Russia \\
  $^{++}$ Landau Institute, Moscow,  Russia  \\
  $^{+++}$Institute of Molecular Biology,Moscow, Russia
\end{center}

\parskip=5mm

\noindent
ABSTRACT

\noindent
We calculate the spectrum of torsional vibrations of a double-stranded
structure that models the double helix of the DNA. We come to the
conclusion that within the framework of the model elementary exictations
may display an asymmetry as regards their winding and direction of the
propagation, depending on initial polarization. The asymmetry could have
a bearing on processes that take place in molecules of the DNA.
\pagebreak

\noindent
Conformational excitations of the DNA molecule are important for
understanding transformations of DNA, for example during
transcription, and therefore there is an urgent need for their investigation.
But, on the other hand, the latter is a difficult problem, even
on the assumption that the excitations are weak enough  to preserve
the hydrogen bonds between base-pairs, so that the double-stranded structure
is deformed but not disrupted. As to our knowledge, it was Cappelmann and
Beim, {\bf \cite{cap}}, who were the first to attempt such a study. They gave a
model of DNA that accommodates a number of modes of elastic vibrations
with sophisticated spectrum, and the analysis of their bearing on the
actual properties of DNA needs further studying. As usual in such cases,
the most important thing is the wise choice of an initial model, which
determines the final outcome; taking into account the qualitative
nature of the approach, one should employ the minimal number of assumptions
so as to obtain simple characteristics of the dynamics of a molecule.
In our work we aim at studying elastic modes of vibrations of the
double-stranded structure, which are
similar to the phonon modes of crystalline lattices.
We show that the asymmetry of the double helix, due to its having
an orientation of the right or the left screw, may result in the propagation
of pairs of waves travelling in opposite directions and producing twists of
different signs in the double-stranded structure.

\noindent
According to the results of {\bf \cite{l1}, \cite{l2}}, the number of open base
pairs at room temperature is very small, so that the system of H-bonds may
be considered intact, within a good approximation. On the other hand the
H-bonds are more fragile than the sugar-phosphate frame, and therefore it
appears to be a reasonable proposition to consider excitations for either
of them separately. Thus, at least for sufficiently small excitational
energies, we may assume the conformation of the sugar-phophate frame being
static, while the H-bonds subject to small deformations.
Hence, the minimal list of basic properties of the DNA
molecule under the circumstances may be drawn as follows:

\begin{itemize}
 \item a molecule of DNA is an elastic system at the spatial scale of the
			 order $500 \AA$ , or less;
 \item the discrete nature of the system is to be taken into account at the
			 intra base-pair distances, that is $3.4 \AA$;
 \item the conformation of the double-stranded structure
			 of DNA may be described, qualitatively, by the vector field
			 $\vec Y$ defined at points of the central line of a molecule,
			 with $\vec Y = 0$ corresponding to the unperturbed
			 state, (cf. {\bf \cite{prd}});
 \item the double helix of the two strands means that the system has a
			 twisted ground state characterized by the twist vector $\vec \Omega $.
\end{itemize}

\noindent
Consequently, the dynamics of the double helix is to be determined by
the twist, and therefore the vector $\vec \Omega$.

\noindent
In accord with the requirements formulated above we may cast the energy
of a molecule in the form

$$
	E = \siN \frac{1}{2M} ( \partial_t \vec P_i)^2 \quad  + \quad
			\siN \frac{K}{2}   (\nabla \vec Y_i)^2  \quad  + \quad
			\siN \frac{L}{2}   \vec Y_i ^2
$$

\noindent
where $\vec Y_i$ is the value of $\vec Y$ at the site corresponding to
the i-th base-pair, and the covariant derivative  $\nabla \vec Y_i$ is given
by the equation

\noindent
$$
	\nabla \vec Y_i =  \frac{1}{a} ( \vec Y_{i+1}  - \vec Y_i) \quad + \quad
								\frac{1}{a} \hat \Omega_i \vec Y_i
$$

\noindent
in which $a$ is the distance between adjacent base-pairs and
the matrix $\hat \Omega_i$ is determined  by the formula

$$
	 \hat \Omega_i = \hat R_i^{-1} ( \hat R_{i+1} - \hat R_i )
$$

\noindent
Here the matrix $\hat R_i$ is the matrix of a rotation
that transforms the local
coordinates at the site $i$ towards the laboratory one. It is important that
the matrix $\hat \Omega_i$ corresponds to
the twist of the double strands
of the helix, and one can describe it by the vector $\vec \Omega$, mentioned
above.  One can derive the dynamical equations for the field $\vec Y$
from the energy given above as usual. The set of values for the matrix
$\hat \Omega_i$ is supposed to be fixed by the equilibrium ground state of the
double helix. It is worth noting that, to a certain extent, the distribution
of base-pairs of different shapes can be mimicked by the choice
of $\hat \Omega_i$.

\noindent
Assuming that the matrices $\hat \Omega_i$ do not depend on the site $i$,
which means that some spatial homogeneity could be accepted,
we are looking for eigenmodes that can be cast in the form

$$
		\vec Y_n = e^{-inqa} Y_q
$$

\noindent
and obtain the modes having the spectrum

\begin{eqnarray*}
  \omega_{1,2}^2  &=&  v^2  \lll  { \dspls 4 A^2 \sin^2 \LL \frac{aq}{2} \RR
			 \pm 2 \sin \phi  \sin(aq)
			 +
			 4 \sin^2 \phi + \Delta^2
			 }  \rr \\
			\omega_3^2 &=&   4 v^2   \sin^2 \LL \frac{aq}{2} \RR   \\
			 v  &=&   \sqrt{ {\dspls \frac{K}{Ma^2} }   }
\end{eqnarray*}

\noindent
in which $A^2$ is given by the equation

$$
  A^2 =  1 + 2 \sin^2 \LL \frac{\phi}{2} \RR
$$

\noindent
the angle $\phi$ is that of the rotation matrix $\hat \Omega$.
The spectrum is illustrated in Fig.1. It is worth noting that
a similar form of the spectrum is obtained for the linearized
expression for matrices $\hat \Omega_i$, that is the rotations
$$
	  \hat R_i^{-1} \cdot  \hat R_{i+1}
$$
being considered infinitesimal (see {\bf \cite{hbonds}} ).
The agreement between the two approaches is in accord
with the qualitative nature of the model we consider.

\noindent
One can infer an important conclusion  from the shape of the spectrum .

\noindent
Let us consider the first mode $\omega_1$ which corresponds to the
minus sign in the equation given above.  The frequency $\omega_1$
has a pronounced minima at $q_*$, so that there are two sets of
waves travelling in opposite directions of the molecule. It should be
noted that the double helix has an orientation, the left or the right
one, due to its screw structure. Thus, we may talk about the sign of
the propagation of a wave. The value of $\omega_1$ at $q_*$ is not zero,
and it results in a winding motion of the vector $\vec Y$ with an
orientation determined by the sign of the propagation, so that for
a pair of waves belonging to the eigenmode $\omega_1$ the windings must be
of different orientaion. We see that the effect depend on the polarization
of a wave, in fact we have considered an eigenmode.
In case we have a mixture of them,  the asymmetry of the winding
can be blurred or totally destroyed.

\noindent
One may cast these arguments in a more quantitative form.  Let us consider
the first mode; its spectrum can be roughly approximated by the expression
$$
	\omega = c | q - q_0| + \omega_0
$$
\noindent
or
$$
	\omega =  \left \{ \begin{array}{ll}
	c q - \omega_1    & \mbox{if $q > q_0$} \\
	- c q + \omega_2  & \mbox{otherwise}
			\end{array}
				 \right.
$$
\noindent
(see Fig.2), in which the frequencies $\omega_1, \omega_2$ read
$$
	 \omega_1 = \omega_0 - c q_0, \quad \omega_2 = \omega_0 + c q_0
$$
It is important that for the numerical values involved, that is
corresponding to the DNA data, the frequencies $\omega_1, \omega_2$ are
positive, as is seen in Fig.2.

\noindent
If we are looking for a solution with
the help of the Fourier transform, we may write its expression in the form
of the decomposition
$$
	u(x,t) = \int^{\infty}_{-\infty} dq \, u_q(t)
$$
in which $u_q(t)$ reads
$$
	u_q(t) = u^0_q \, e^{i(\omega t - q x)}
$$
and $\omega$ depends on $q$. It is easy to convince oneself that the
solution can be cast in the form
$$
	u(x,t) = \int^{q_0}_{-\infty} dq \,
													\dspls{ [ u^0_q \, e^{-i(t+x)q}]} e^{i\omega_2 t}
				 + \int^{\infty}_{q_0}  dq \,
													\dspls{ [ u^0_q \, e^{-i(t-x)q}]} e^{-i\omega_1 t}
$$
or, otherwise
$$
	u(x,t) = u_+(t+x) \, e^{\omega_2 t} + u_-(t-x) \, e^{-\omega_1 t}
$$
The latter desribes the two waves travelling in opposite directions and
having different windings.

\noindent
We made a numerical simulation of the dynamics of our model.
The asymmetry is seen in that two waves travelling from an initial
disturbance in opposite direction may oscilate having equal phases of
different sign.
As is expected, the effect of the asymmetry depends
on an initial polarization, which corresponds to a superposition
of the eigenmodes indicated above, so that for a certain direction
it can be totally suppressed, whereas for those belonging to
the eigenmodes it is pronounced.

\noindent
On considering the estimates for the values of the density
per unit of length, approximately $ 10^{-14} gr/cm $, and
according to the papers {\bf \cite{Strick}, \cite{const} },the elastic
constant,  $\sim 10^{-6} erg/cm$,
me may suggest, therefore, that the velocity of the waves considered
above is of the order $10^5 cm/sec$. Consequently, the frequencies of the
modes, even for the region of wavenumbers corresponding to the dips of the
spectrum seen in Fig.1, are those of hypersound, that is $10^{11}$ or
$10^{12}$ or even more, depending on the values of the elastic constant.
The values are by many orders of magnitude above those involved in
biological process. Their relation to the latter presents a very
interesting problem. In fact, we shall investigate the influence of
super-high frequency dynamics on the very low frequncy one. It should be
noted that so far no electronic properties have been taken into account
in conjunction with the elastic ones, and in this respect the approach
of elasticity theory, generally employed now, might be incomplete. Studying
the elastic excitattions in the DNA could be a means for finding the
answer.

\noindent
Another important question is the influence of ambient media that
should result in a very strong dissipation of excitations.
Whether they will be overdamped, and to all practical purposes
unobservable , as appears to be the case in a similar situation in
proteins, remains to be seen. A strong argument in favour of their playing
an important part in the dynamics of the DNA, is the regidity of
the latter displayed by its large persistence length.

\noindent
At any rate, the investigation of elementary excitations
should be a necessary
prerequisite for investigating conformational changes
due to the action of
external agents, e.g. enzymes. The salient feature is that in these
circumstances a preferential direction of the promoter's action is often
observed {\bf \cite{liu} }, as well as a torsional dynamics of the
double helix. Thus, the anysotropy of the propagation of a disturbance
due to the screw orientation of the double helix
for which we make out a case in this paper, appears to be very
interesting and worth further studying.

\noindent
This research was supported by RFFR.
One of the authors (VLG) is thankful
to CNRS for the financial support, and ENS-Lyon for the hospitality.
The authors are thankful to Michel Peyrard for many fruitful
discussions and comments.

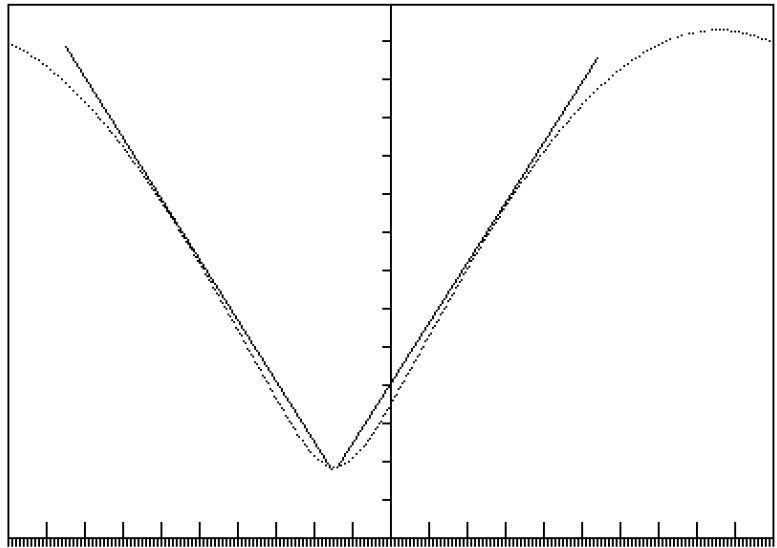
\end{document}